\documentclass[12pt,preprint]{aastex}
\usepackage{psfig}
\def\HI {H\kern0.1em{\sc i}} 
\def\radm {rad m$^{-2}$} 

\def\dg{^{\circ}}

\begin{document}
\title{~~\\ ~~\\ Faraday Rotation Measures in the parsec scale jets \\
of the radio galaxies M87, 3C\,111, and 3C\,120}
\shorttitle{Jet RMs in M87, 3C\,111, \& 3C\,120}
\shortauthors{Zavala \& Taylor}
\author{R. T. Zavala\altaffilmark{1,2} \& G. B. Taylor\altaffilmark{1}}
\email{rzavala@nrao.edu, gtaylor@nrao.edu}
\altaffiltext{1}{National Radio Astronomy Observatory, P.O. Box 0, Socorro, NM 87801}
\altaffiltext{2}{Department of Astronomy, New Mexico State University, MSC 4500 P.O. Box
30001, Las Cruces, NM 88003-8001}


\slugcomment{Accepted to the Astrophysical Journal Letters}

\begin{abstract}
Parsec scale Faraday rotation measure maps are presented for the radio galaxies
M87, 3C\,111, and 3C\,120. These VLBA observations were made at 8, 12, and 15 GHz. 
M87 has an extreme RM distribution which varies from 
$-$4000 \radm\ to more than 9000 \radm\ across a projected distance of
0.3 parsecs in its jet. M87 has no polarized flux closer than 17 mas from 
the core. 3C\,111 and 3C\,120 both show polarized emission in their cores
which is consistent with the expectations of unified schemes for these 
broad line radio galaxies. 3C\,111 has an RM gradient which increases from 
$\sim -200$ \radm\ 4 mas from the core to $\sim -750$ \radm\ on the side of the 
jet closest to the core. 3C\,120 has a more moderate RM distribution in the jet
of approximately 100 \radm\ but this increases by an order of magnitude in the core.

\end{abstract}


\section{Introduction}

Several recent papers (Udomprasert et al. 1997, Cotton 1997 and Taylor 1998 \&
2000) have demonstrated that extreme 
values of Faraday rotation of up to 40000 \radm\  in the rest 
frame of quasars are possible. \citet{zt01} showed that the Rotation Measure (RM)
properties of quasars vary on both small spatial (parsec) scales and short
timescales (1.5$-$3 years). These observations suggest that the observed RM distributions are 
intrinsic to the central few hundred parsecs of AGN and are not a foreground effect 
from the host galaxy ISM, the ICM, or the Milky Way. Hence, the observed Faraday rotation 
can serve as a probe of the magnetic field weighted by the electron density along the line 
of sight near the central engines of AGN. With the electron density supplied by spectral line 
diagnostics (e.g. Osterbrock 1989) an estimate of the magnetic field strength and orientation
can be made provided one assumes a physically reasonable path length.

To date parsec-scale RM observations have only been made for quasars. Polarimetric observations 
of other classes of AGN are required in order to test the unified model for AGN \citep{ant93} 
by looking for orientation effects in the RM distributions. In order to form a more complete 
picture of the parsec scale RM properties of AGN \citet{tay00} designed a sample of 40 quasars, 
BL Lacs, and radio galaxies for polarimetric observations. We have completed this survey  
with the VLBA and present here the first sub-parsec scale RM maps for radio galaxies. 
 
We assume H$_0 = 50$ km s$^{-1}$ Mpc$^{-1}$ and q$_0$=0.5 throughout.
  
\section{Observations and Data Reduction}
The observations, performed on 2000 June 27 (2000.40), were carried out at 
seven widely separated frequencies between 8.1 and 15.2 GHz using the 10 element VLBA\footnote
{The National Radio Astronomy Observatory is operated by 
Associated Universities, Inc., under cooperative agreement with the National 
Science Foundation.}. Amplitude calibration and fringe-fitting were performed as in \citet{tay00}. 
Feed polarizations of the antennas were determined using the AIPS task
LPCAL. This calibration was performed twice, first using J0136$+$478 with 
an accompanying CLEAN model for this polarized source. Plots of the real versus imaginary 
crosshand polarization data indicated that a satisfactory d-term solution was obtained. After
this calibration was performed J2022$+$616 was found to be unpolarized. The d-terms were solved 
for again using J2022$+$616 as an unpolarized calibrator as a consistency check. Both 
sources produced nearly identical corrections for the d-terms, and J2022$+$616 was used as the
final d-term calibrator.   

Absolute electric vector position angle (EVPA)
calibration was determined by using the EVPA's of J0854$+$201, J0927$+$390,
J1310$+$323 and J2202$+$422 listed in the VLA Monitoring 
Program\footnote{http://www.aoc.nrao.edu/$\sim$smyers/calibration/} \citep{tmy00}. 
The EVPA calibration at 8 GHz was readily
obtained from the polarization calibration website. The  observations
were interpolated in 
time as necessary. Polarization monitoring observations at 8 and 22 GHz were interpolated 
to produce position angles at 12 and 15 GHz. Figure 1 shows calibrated 
position angle results for these four calibrators with the associated
VLA data from the website. As a further test of our EVPA calibration 
we examined the rotation measure of component C4 of 3C\,279 after the EVPA calibration 
was completed. In \citet{zt01} C4 was shown to have a relatively low RM and $\chi = 
-84^{\dg}$ at 8 GHz to $-87^{\dg}$ at 15 GHz 2000 January 27. After
the EVPA calibration for this experiment was completed we measured C4 and the EVPA
was $-87\dg$\  for all frequencies. Although C4
does vary its position angle and RM with time the agreement between the observations of 
2000 January and 2000 July suggests that the polarization calibration was performed correctly.  

To perform the rotation measure analysis data cubes in $\lambda^2$ were constructed. There
are four points across the 8 GHz band, and three points at 12.1, 12.5, and 15.17 GHz. 
This provides adequate short and long spacings in $\lambda^2$ to properly
recover RMs between $\pm$ 30000 \radm. The 12 and 15 GHz 
images used to produce the polarization angle maps were tapered to approximate the 8 GHz 
resolution, and a restoring beam matched to the 8 GHz beam was used. 

\section{M87 Results}

This well studied galaxy was shown by \citet{jun01} to have an average RM of $-4400$ \radm\  
which varied from $-2000$ to $-12000$ \radm. Our observations similarly show an extreme 
RM of $-4831 \pm 140$ to $+9563 \pm 265$ \radm\ 22$-$24 mas from the core (Fig. 2). This 
RM distribution is coincident with a peak in total intensity which appears displaced 
northward from the inner 15 mas of the jet. The RMs are mostly between $-3000$ and $-5000$ 
with the exception of a patch where the RMs increase in magnitude by a factor two while 
changing sign to a positive slope. The sign change of the slope is significant, as it 
requires the magnetic field to change direction over a projected distance of 0.3 pc. Good 
agreement to the $\lambda^2$ signature of Faraday rotation is seen as indicated by the
inset plots in Figure 2.

Although the significant polarized flux at 8 GHz is limited
to the region of the RM signal there is some (4.5 $\pm$ 0.4 mJy) polarized emission at 15 GHz
closer to the core. This occurs at $\sim$ 17 mas from the core, in a ${\it valley}$
of total intensity (at 15 GHz) between the inner jet and the displaced peak in I at 22$-$24 mas.  
The lack of corresponding polarized emission at 8 or 12 GHz is consistent with Faraday
depolarization at lower frequencies if the RM gradient is greater than 
2370 \radm\ mas$^{-1}$.  

\section{3C\,111 Results}
This broad line radio galaxy (BLRG) \citep{sar77} exhibits a gradient in RM towards the core. The 
RM changes from $\sim -212 \pm 66$ \radm\ at 4 mas from the core to $\sim -730 \pm 78$ \radm\ on 
the side of the jet closest to the core. As for M87, the polarized emission at 8 GHz for 
3C\,111 is limited to the region of the RM color overlay in Figure 3. However, significant 
polarized emission at 15 GHz (5.9 $\pm$ 0.4 mJy) exists within 2 mas 
of the core in 3C\,111. At the location of the left-hand insert in Figure 3 the percent
polarization is 12$\%$ at 15 GHz. The percent polarization 1.3 mas NE of the core is 
1.5$\%$ at the same frequency.

\section{3C\,120 Results}
3C\,120 is another BLRG \citep{tad93} and its polarization properties were monitored 
over 16 months with the VLBA by \citet{gmajg00}. \citet{gmajg00} reported an RM in the jet 
of 3C\,120 of $6000 \pm 2400$ \radm\ based on measurements at 22 and 43 GHz. Our observation
a little more than a year after G{\' o}mez et al. shows a much smaller RM distribution. In 
Figure 4 it is apparent  that 3C\,120 has the lowest jet RM distribution of the three radio 
galaxies presented here, $105 \pm 72$ \radm. The time variability of parsec scale 
RMs as seen in \citet{zt01} is also present in 3C\,120 \citep{gmajg00} and may account for 
the difference between our results and those of G{\' o}mez et al.  

Contrary to M87 and 3C\,111, 3C\,120 has detectable polarized emission at all frequencies 
in the core and RMs of $|1000 - 2000|$ \radm\ are present. As seen in Figure 4 the RM fits 
show considerable scatter at 8 GHz, and the reality of the sign change in the core RM 
distribution (from $+$2000 to $-$1000 \radm) is suspect as the negative slope area shows 
deviations from a $\lambda^2$ law. The polarized flux within the core ranges from 4 $\pm$ 
0.6 mJy at 8 GHz to $\sim 30$ $\pm$ 0.6 mJy at 12 and 15 GHz. The range of $m$ (15GHz) in
the jet reach maxima of 9$-$16$\%$, but within 1 mas from the center of the Stokes I image 
$m$ falls to 3$\%$ or less. 

\section{Implications for Unified Schemes}
Unified models for AGN describe radio galaxies as substantially inclined to the line of sight.
This scenario places the core of a radio galaxy behind a dense multi-phase obscuring disk. 
The disk is expected to produce a high enough RM that the radio galaxy core is essentially 
unpolarized. Such a picture appears reasonable for M87, which is classified as a narrow line 
radio galaxy \citep{sspin79}. This picture is supported by 
the lack of any polarized emission from the core of M87 (Figure 5). The requisite disk was first 
reported by \citet{harms94}, although recent IR observations of M87 failed to detect the expected
dust torus \citep{perl01}. Even if the thick dust torus is absent a mechanism which accounts for 
the lack of polarized flux less than 17 mas from the core is still required. Invoking a 
depolarizing Faraday screen requires such a screen to have a thickness of at least 3 parsecs. 
The RM data also require that the magnetic field be 
disordered on sub-parsec scales due to the sign change seen in the RM in Figure 2.

3C\,111 and 3C\,120 are classified as BLRGs and thus are expected to have 
jets more inclined towards the observer. Therefore, more polarized emission near the core might
be expected as the central engine is uncovered by the obscuring torus. This is supported by
the presence of polarized flux density at 15 GHz in the core for 3C\,111 and at all three 
frequencies for 3C\,120, consistent with their classification as BLRGs. Figure 5 illustrates 
this argument as both 3C\,111 and 3C\,120 have polarized emission at 15 GHz in their central 
regions.  

Some caveats must be presented in this interpretation. Although M87 is classified as a NLRG,
\citet{ssc99} report that the Ly$\alpha$ emission line has a width of 3000 km sec$^{-1}$ 
which suggests that at least some of the broad line region (BLR) is indeed visible. Using a black hole 
mass of 2.4 $\times 10^9$ M$_\odot$ \citep{harms94} a velocity of 1860  km sec$^{-1}$ 
is achieved at an orbital radius of 3 parsecs. Thus, the RM distribution observed in M87 may be
produced by the BLR itself. M87 is also a relatively low luminosity radio galaxy 
and may not provide a realistic comparison to more luminous objects. Lastly, the proximity of M87 
also serves to expose details in its inner structure which may escape notice in 3C\,111 and 
3C\,120 due to their greater distance. 
   
\section{Nature of the Faraday Screen}
Given that Faraday rotation occurs, what part of the AGN plays the role of the Faraday
screen? In \citet{tay00} this role was assigned to the narrow line region (NLR) itself, 
and the assumed properties of the NLR seemed consistent with the observed RMs. However, the NLR
is assumed to have a low volume filling factor $\epsilon$ \citep{sspin79}.
Taking a typical value for $\epsilon$ of
$10^{-3}$ gives a covering factor of 0.01. Therefore, assuming a random distribution of NLR 
clouds, one should not expect the coherent RM distributions observed if only 1 percent of the 
beam is intersected by the screen. Alternatively, the screen may consist of clouds which are 
entrained by the jet, perhaps in a boundary layer. This would account for the observed covering 
factor of the screen. A similar idea is put forth by \citet{gmajg00} who propose a cloud 
0.4 pc across to explain the light curves and polarization properties of components in 3C\,120. 

To estimate a B field for a boundary layer screen in M87 we assume a scale size of 
0.3 parsecs for clumps within the screen to match the scale size of the sign change of the RM 
slope. We use the electron density n$_e$ of 1100 cm$^{-3}$ from \citet{sspin79} to obtain a magnetic 
field strength of 16$-$34 $\mu$Gauss. Such a field strength seems weak compared to Zeeman 
measurements in the center of our own galaxy of 0.5 mG \citep{cru96} to 3 mG \citep{pla95}. 
If the magnetic field is in equilibrium with a gas at a temperature of 10$^4$K and using the 
electron density of \citet{sspin79} a magnetic field strength of 200$\mu$Gauss is obtained.

The presence of broad Ly$\alpha$ lines in M87 prompts consideration of the BLR as a 
Faraday screen. As discussed in \S 4 the observed Ly$\alpha$ velocity could be found within 3 
parsecs of the black hole, placing the Ly$\alpha$ clouds along the line of sight to the 
jet in our VLBA images.  However, at the expected density of n$_e$ 10$^{10}$ 
cm$^{-3}$ \citep{mc85} the BLR clouds are optically thick at 8$-$15 GHz \citep{mar75}.  

Lastly, we consider the hot, tenuous gas that is expected to confine the NLR clouds. 
Such a gas is expected to be disordered and radiation propagating through it should 
experience a random walk RM effect. This would produce an RM with a zero mean, Gaussian
distribution. Such a distribution is definitely lacking for M87 and 3C\,111, and possibly
for 3C\,120 as well. Without tighter constraints on the physical parameters it is not possible
to discriminate between these, or other, Faraday screen scenarios.

\acknowledgments
We thank Avi Loeb for pointing out the possible significance of the BLR in
the RM distribution of M87. We also thank J. L. G\'{o}mez for comments on the
manuscript. This research has made use of the NASA/IPAC Extragalactic 
Database (NED) which is operated by the Jet Propulsion Laboratory, Caltech, under
contract with NASA, and NASA's Astrophysics Data System Abstract Service.

\clearpage

\begin{figure}
\plotone{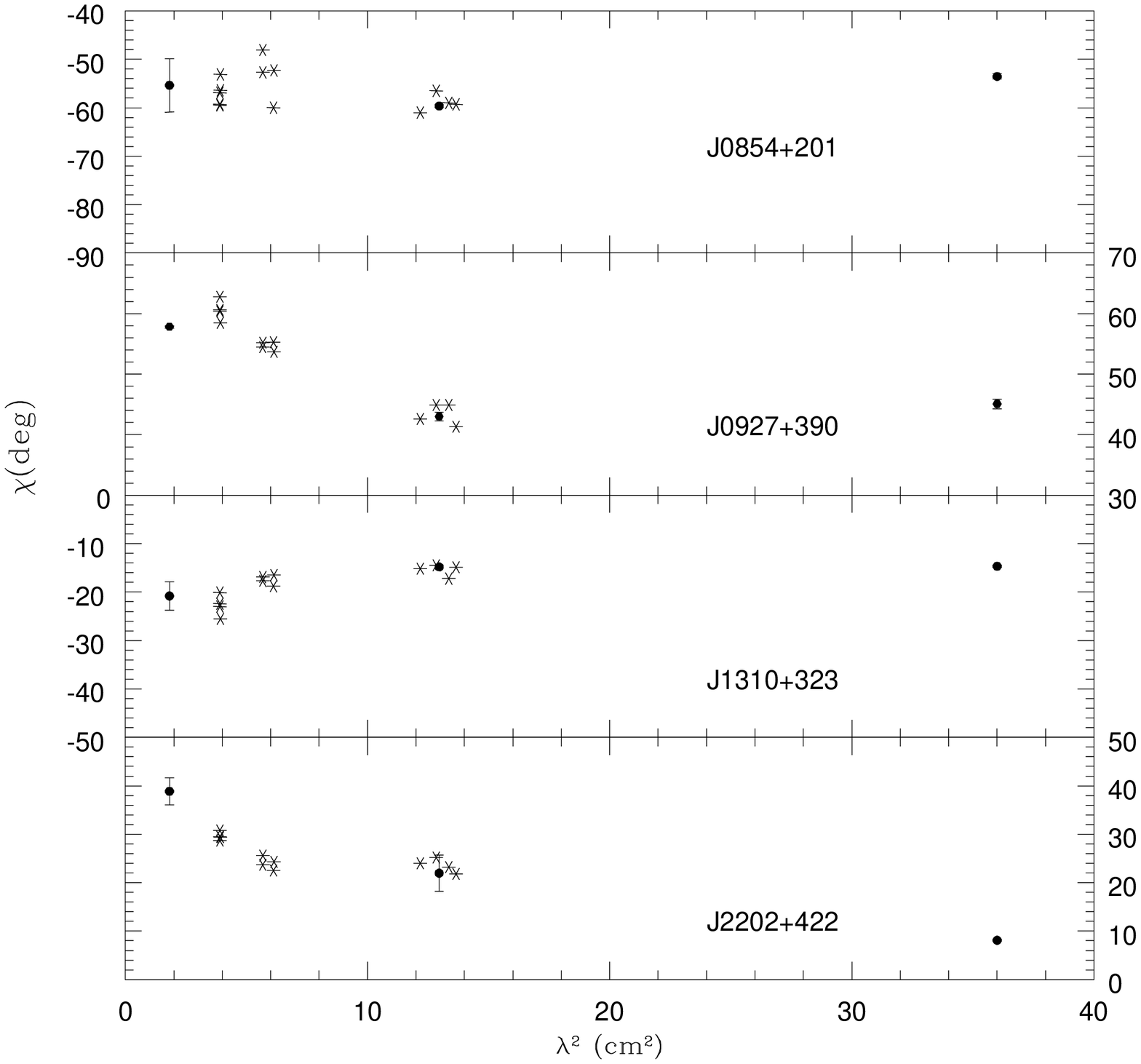}
\caption{Plot of EVPA vs. $\lambda^2$ for calibrators listed in \S 2.  Asterisks are 
EVPA after applying correction based on VLA polarization monitoring data. Filled circles
represent VLA polarization monitoring data for the calibrators. 
Uncertainty in position angle is estimated as ${\pm 2^{\rm o}}$ for 
8 GHz and ${\pm 4^{\rm o}}$ for 12 and 15 GHz after EVPA calibration. \label{fig1}}
\end{figure}
\clearpage

\begin{figure}
\epsscale{0.8}
\plotone{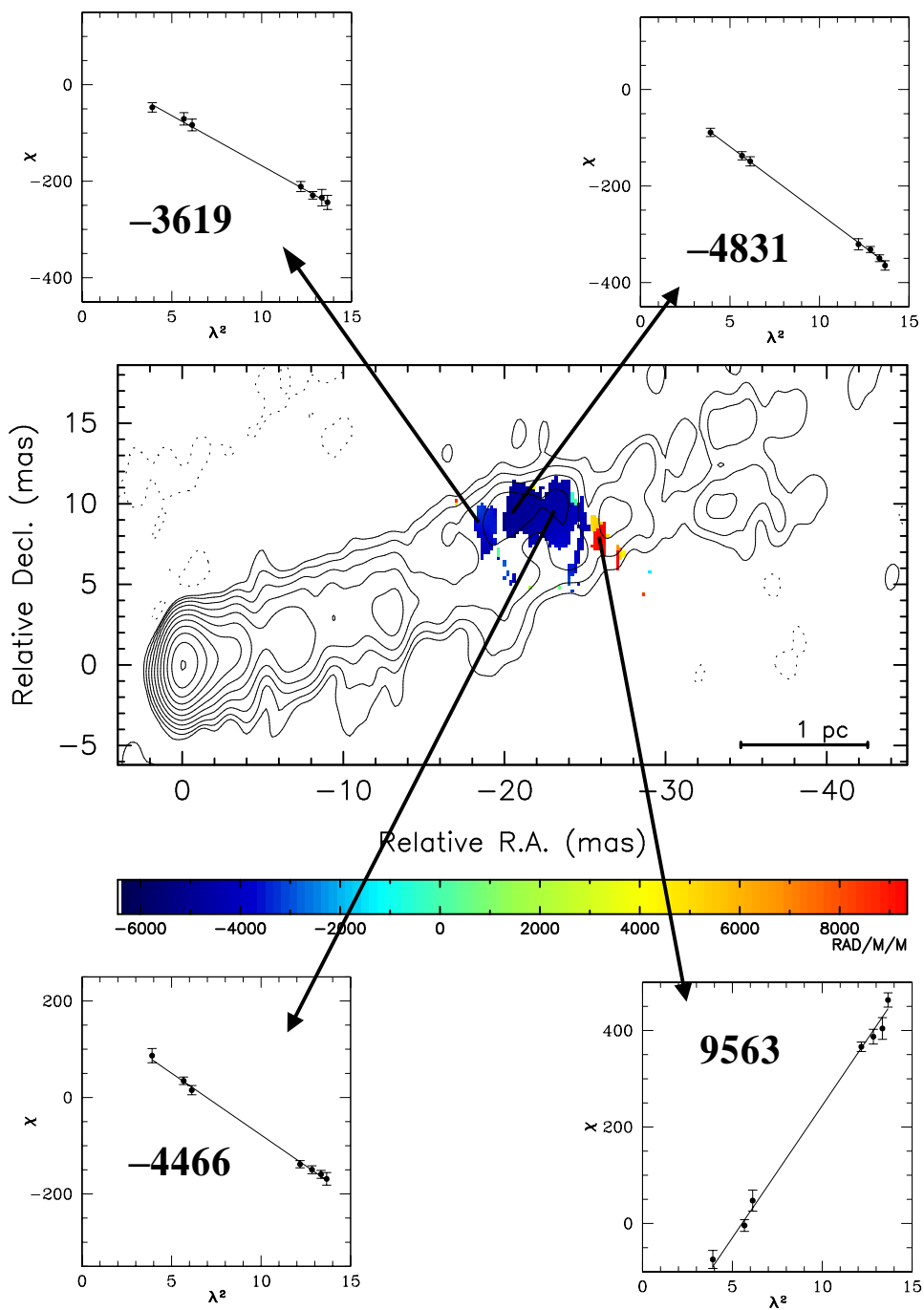}
\caption{RM image of M87 from data at 8 to 15 GHz,
with contours of total intensity at 8 GHz overlaid.  The 
restoring beam has dimensions 1.0 $\times$ 2.7 milliarcsec at position angle 
0$\dg$. The colorbar ranges from $-$6300 to $+$9300 \radm\ .
Contours start at 1 mJy/beam and increase by factors of 2. \label{fig2}}
\end{figure}
\clearpage

\begin{figure}
\epsscale{0.8}
\plotone{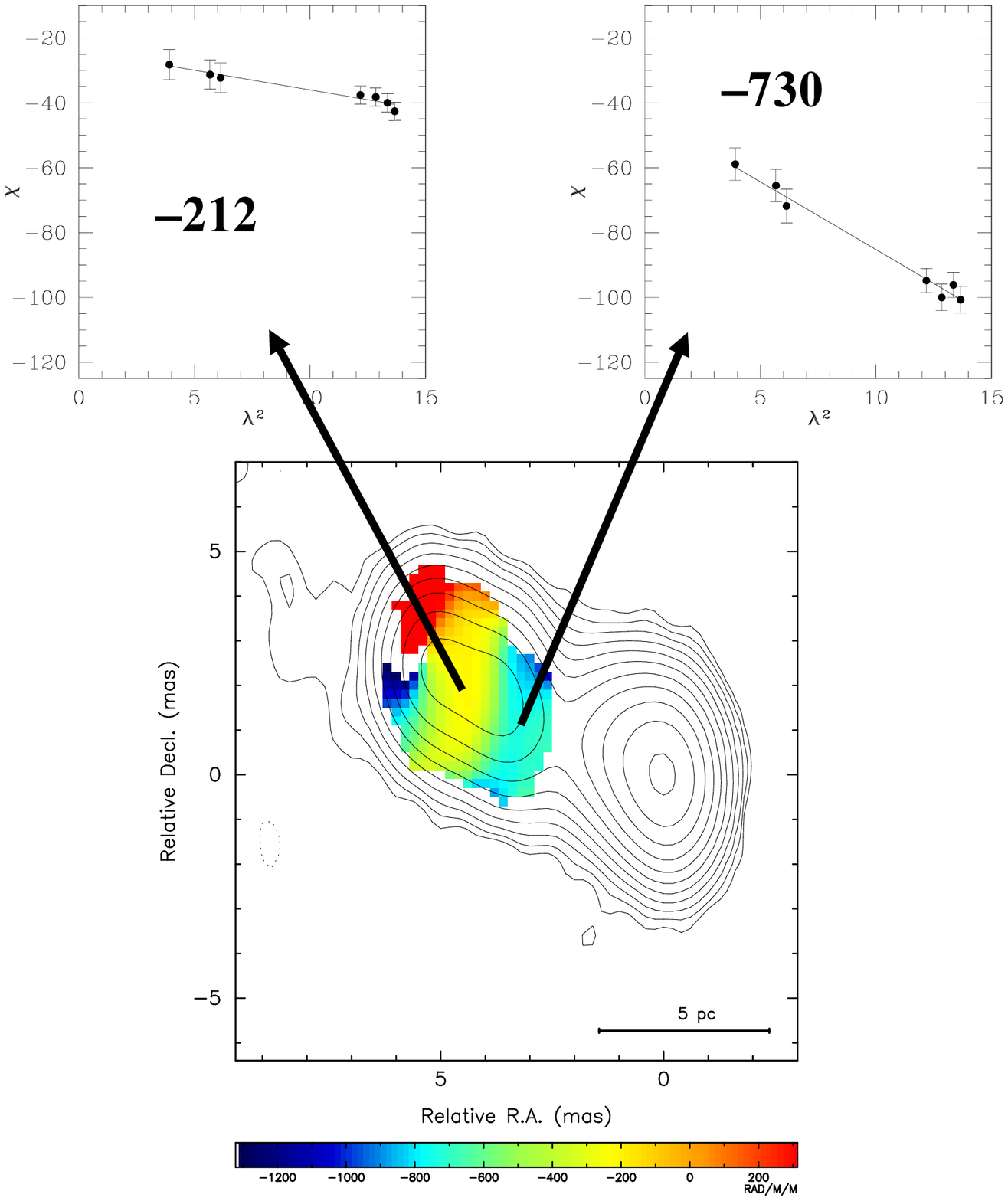}
\caption{RM image of 3C\,111 from data at 8 to 15 GHz,
with contours of total intensity at 8 GHz overlaid. The insets display 
rotation measure fits at the points indicated in \radm. The restoring beam has
dimensions of 1.1 $\times$ 2.1 milliarcsec at position angle 0$\dg$. The 
colorbar ranges from $-$1300 to $+$300 \radm. Contours start at 0.75 mJy/beam 
and increase by factors of 2. \label{fig3}}
\end{figure}
\clearpage

\begin{figure}
\plotone{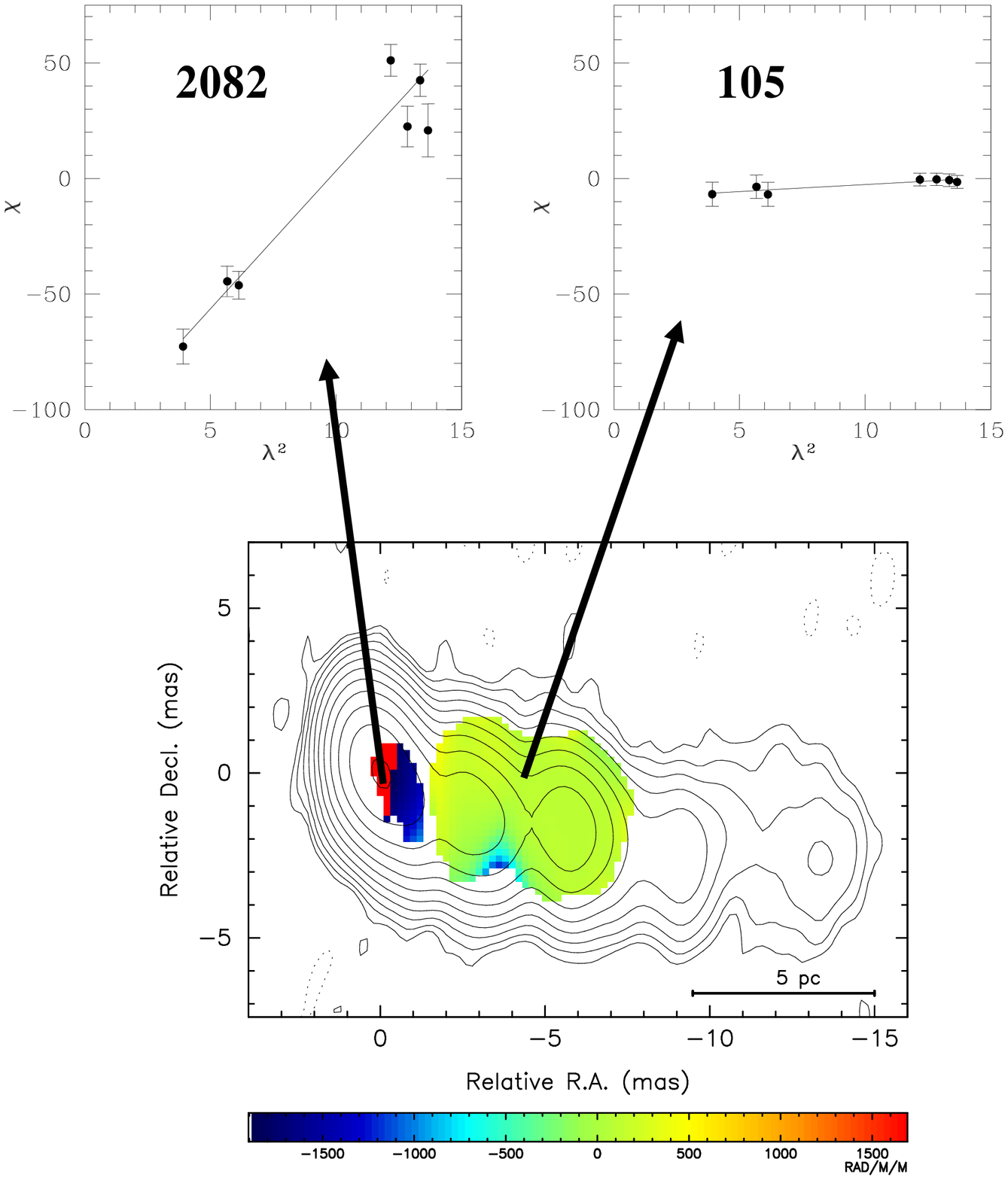}
\caption{RM image of 3C\,120 from data at 8 to 15 GHz,
with contours of total intensity at 8 GHz overlaid. The insets display 
rotation measure fits at the points indicated in \radm. The restoring beam has 
dimensions 1.1 $\times$ 2.6 milliarcsec at position angle 0$\dg$. The 
colorbar ranges from $-$1900 to $+$1700 \radm. Contours start at 1 mJy/beam and 
increase by factors of 2. \label{fig4}}
\end{figure}
\clearpage

\begin{figure}
\plotone{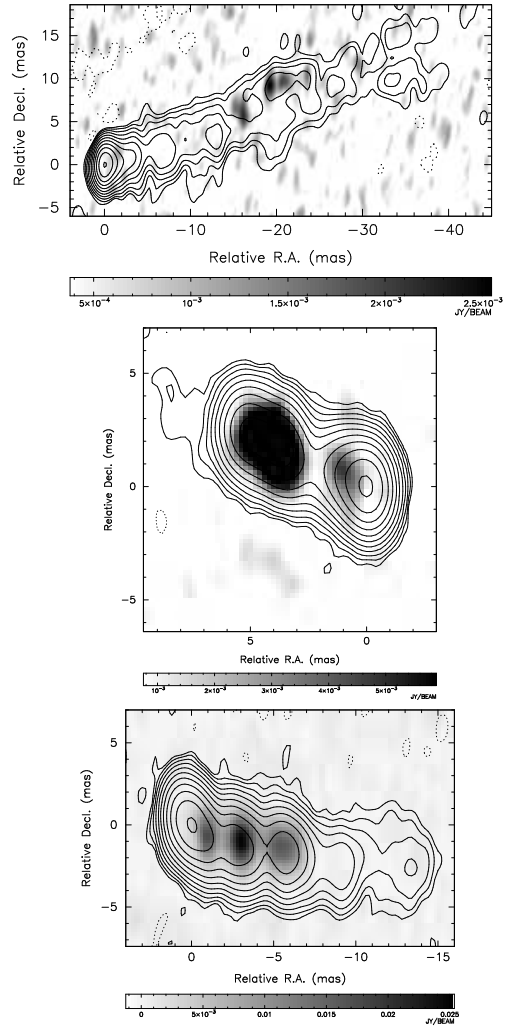}
\caption{Mosaic of 8 GHz total intensity (contours) with 15 GHz 
polarized intensity (greyscale) overlaid. The contours are the same as
for Figures 2-5. \label{fig5}}
\end{figure}
\clearpage

\end{document}